\input epsf
\input rotate
\documentstyle[12pt]{article}
\begin{document}
\newcommand{\linespace}[1]{\protect\renewcommand{\baselinestretch}{#1}
  \footnotesize\normalsize}
\newcommand{\alfs}{\mbox{$\alpha_s$}}
\newcommand{\asop}{\mbox{$\frac{\alpha_s}{\pi}$}}
\newcommand{\qsq}{\mbox{$Q^2$}}
\newcommand{\qnsq}{\mbox{$Q_0^2$}}
\newcommand{\mztwo}{\mbox{$M_Z^2$}}
\newcommand{\glsint}{\mbox{$\int F_3dx$}}
\newcommand{\lmsb}{\mbox{$\Lambda_{\overline{MS}}$}}
\begin{center} 
\vspace*{4.cm} 
A MEASUREMENT OF $\alpha_s(Q^2)$ 
FROM THE GROSS LLEWELLYN SMITH SUM RULE
\end{center} 
\vspace{24pt} 
\begin{sloppypar}
\centering
\noindent
        D.~A.~Harris$^8$, C.~G.~Arroyo$^4$, P.~Auchincloss$^8$,  
        P.~de~Barbaro$^8$, A.~O.~Bazarko$^4$, R.~H.~Bernstein$^5$, 
        A.~Bodek$^8$, T.~Bolton$^6$,
        H.~Budd$^8$, J.~Conrad$^4$, R.~B.~Drucker$^7$, 
        R.~A.~Johnson$^3$, J.~H.~Kim$^4$, B.~J.~King$^4$, 
        T.~Kinnel$^9$, G.~Koizumi$^5$, S.~Koutsoliotas$^4$,
        M.~J.~Lamm$^5$, W.~C.~Lefmann$^1$, 
        W.~Marsh$^5$, K.~S.~McFarland$^5$, C.~McNulty$^4$, 
        S.~R.~Mishra$^4$, D.~Naples$^5$, P.~Nienaber$^{10}$, 
        M.~Nussbaum$^3$, M.~J.~Oreglia$^2$, 
        L.~Perera$^3$, P.~Z.~Quintas$^4$, A.~Romosan$^4$,
        W.~K.~Sakumoto$^8$, B.~A.~Schumm$^2$,
        F.~J.~Sciulli$^4$, W.~G.~Seligman$^4$, M.~H.~Shaevitz$^4$, 
        W.~H.~Smith$^9$, P.~Spentzouris$^4$,
        R.~Steiner$^1$, E.~G.~Stern$^4$,
        M.~Vakili$^3$, U.~K.~Yang$^8$
%
%
\end{sloppypar}
\vglue 0.2cm
\vspace{.1in}  
\begin{verse}
\baselineskip=13pt 
$^1$ Adelphi University, Garden City, NY 11530 \\
$^2$ University of Chicago, Chicago, IL 60637 \\
$^3$ University of Cincinnati, Cincinnati, OH 45221 \\
$^4$ Columbia University, New York, NY 10027 \\
$^5$ Fermi National Accelerator Laboratory, Batavia, IL 60510 \\
$^6$ Kansas State University, Manhattan, KS 66506 \\
$^7$ University of Oregon, Eugene, OR 97403 \\
$^8$ University of Rochester, Rochester, NY 14627 \\
$^9$ University of Wisconsin, Madison, WI 53706 \\
$^{10}$ Xavier University, Cincinnati, OH 45207 \\
\end{verse}
\begin{center} 
Talk Presented by D.A. Harris 
\end{center} 
\vspace{4.55cm} 
\begin{abstract} 
The Gross Llewellyn Smith sum rule has been measured at different values of 
four-momentum transfer squared ($Q^{2}$) by combining the
precise CCFR neutrino data with data from other deep-inelastic scattering
experiments at lower values of $Q^{2}$.  A comparison with the 
${\cal O}(\alpha^{3}_{s})$ predictions of perturbative QCD   
yields a determination of $\alpha_{s}$ and its dependence on  $Q^{2}$ 
in the range $1\,GeV^2 < Q^{2} < 20 \,GeV^{2}$.  Low \qsq\ tests have 
greater sensitivity to \alfs(\mztwo) than high \qsq\ tests, since at low \qsq\ 
\alfs\ is large and changing rapidly.  
\end{abstract} 

\pagebreak
To leading order in perturbative QCD, the structure function $xF_3$
measured in $\nu$N scattering is the difference between the quark and
anti-quark momentum distributions.  The GLS sum rule predicts that the
integral over $x$ of $F_3$ is simply 3, the number of valence
quarks in a nucleon \cite{gls}.  There are corrections to
the sum rule which introduce a dependence of the GLS integral on
$\alpha_s$, the strong coupling constant, in the following way
\cite{gls_higher_order}:
\begin{equation} 
\int_{0}^{1} xF_3(x,Q^2)\frac{dx}{x} = 
3(1 - \frac{\alpha_s}{\pi} 
- a(n_f)(\frac{\alpha_s}{\pi})^2 
- b(n_f)(\frac{\alpha_s}{\pi})^3 )
- \Delta HT 
\label{eqn:glsalf} 
\end{equation}
where $a$ and $b$ depend on the number of quark flavors, $n_f$, 
accessible at a given $x$ and four-momentum transfer squared, $Q^2$.  
$\Delta HT$ represents a higher twist
contribution, which has been estimated using QCD sum rules, 
a Vector Meson Dominance Model, and a Non-relativistic Quark Model to be  
$0.27\pm0.14/Q^2(GeV^2)$\cite{bktwist}.  
The \qsq\ dependence of $\alpha_s$ is as follows \cite{marciano}:  
\begin{equation}
\frac{\alpha_s(Q^2)}{4\pi} = 
                \frac{1}{\beta_0(n_f)
\ln\left( \frac{Q^2}{\lmsb^2}\right)}  - 
                \frac{\beta_1(n_f)}{\beta_0(n_f)}
\frac{\ln\left(\ln\left(\frac{Q^2}{\lmsb^2}\right)\right)}
{\beta_0(n_f)^2\ln^2\left(\frac{Q^2}{\lmsb^2}\right)}
+ {\huge{\cal O}} 
\left( \frac{1}{\ln^3\left( \frac{Q^2}{\lmsb^2} 
\right)} \right).    
\label{eqn:alfq2} 
\end{equation}

The challenge in evaluating \glsint\  is that for a given \qsq\ 
value, there is a limited $x$ region that is accessible by any one 
experiment.  The incoming neutrino energy imposes a minimum $x$ constraint 
and detector acceptance imposes a maximum $x$ constraint.  
CCFR has data at low \qsq\  and low $x$ ($10^{-2}<x<10^{-1}$), and at high 
\qsq\  and high $x$ ($10^{-1}<x<1$).  
The CCFR detector and the measurement of $xF_3$ have been 
described in detail elsewhere \cite{ccfrboth}.  
One way to evaluate \glsint\ over all $x$ is to extrapolate $xF_3$
from all \qsq\ regions to a \qnsq\ value where the data is predominantly
at low $x$.  A previous CCFR analysis found that for $\qnsq=3\,GeV^2$,
$\int F_3dx = 2.50\pm.018(stat)\pm.078(syst)$\cite{oldgls}.  By using
QCD to extrapolate $xF_3$ to \qnsq\, however, one introduces \alfs\
{\it a priori} into the problem.  Furthermore, higher twist effects
are not included in QCD extrapolations.

The goal of this analysis is to evaluate \glsint\ without introducing 
any {\it ad hoc} \qsq\ dependence.  By combining the CCFR data with 
that of several other experiments enough data at 
different energies are obtained to measure \glsint\ without 
\qsq\ extrapolation at values of 
\qsq\ between $1\,GeV^2$ and $20\,GeV^2$.  
The $xF_3$ measurements from experiments WA59, WA25, 
SKAT, FNAL-E180 \cite{allxf3}, and
BEBC-Gargamelle \cite{bebcg} were normalized to the CCFR $xF_3$ measurements
in the \qsq\ regions of overlap and then were  
used along with the CCFR $xF_3$
data.  Furthermore, since at high $x$ the structure function
$F_2\approx xF_3$, one can use $F_2$ data from $e^-$N scattering at
SLAC \cite{slac} in this region ($x>0.5$) by normalizing it to the ratio of
$xF_3/F_2$ as measured in the CCFR data.  This is particularly
important at low \qsq\ where there is no $xF_3$ data at high
$x$.  The published CCFR $xF_3$ data were modified for new electroweak
radiative corrections (Bardin\cite{bardin}).  In addition, the CCFR data
were corrected for the contribution from the strange sea \cite{bazarko} of 
events containing two oppositely charged muons.  
Finally, by comparing the $F_2$ values
of CCFR to those from SLAC \cite{slac}, NMC 
and BCDMS \cite{twof2}, the overall normalization of the CCFR data 
was determined to be $1.019\pm0.011$.
\begin{figure}[t]
    \begin{center}
    \leavevmode   \epsfxsize=17.cm
    \epsfbox{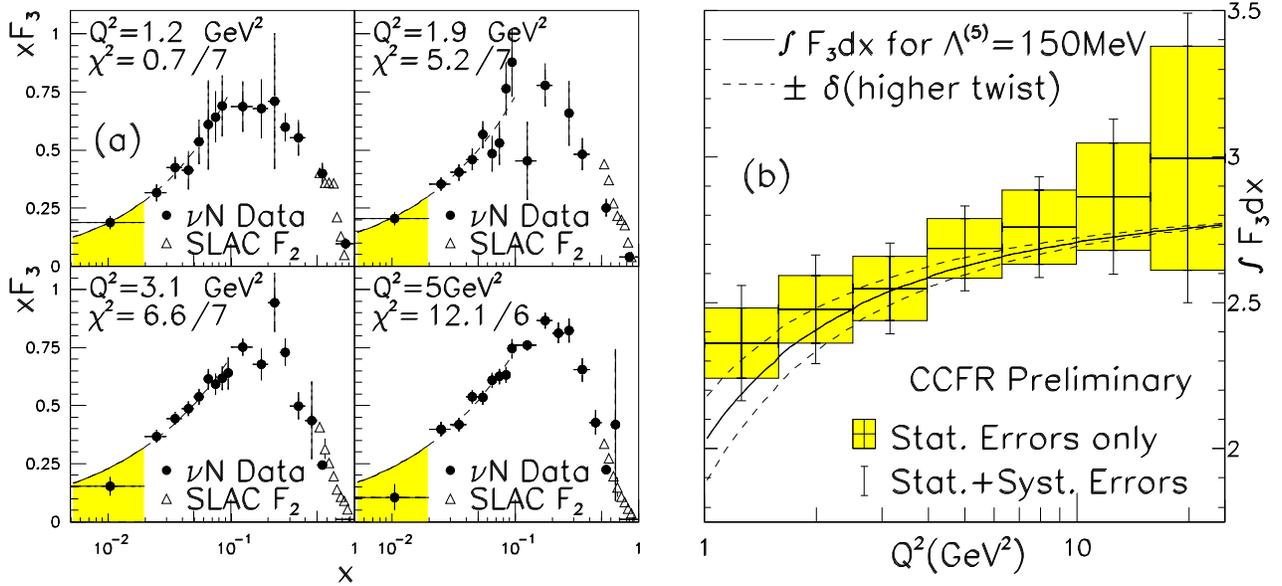}
    \end{center}
    \vspace{-.4in}  
    \caption{(a)$xF_3$ vs. $x$ for four different low 
             \qsq\ values. The function shown is a power law fit to data  
             below $x=0.1$.
             (b)$\int F_3dx$ vs. \qsq, and the theoretical
             prediction for the integral for $\lmsb^{(5)}=150\,MeV$.
             The dashed lines represent the uncertainty in 
             the higher twist correction.}
    \label{fig:combine}
\end{figure}
To integrate over all $x$, this analysis sums the binned data for
$x>0.02$.  For the contribution to the integral at lower $x$, the data
below $x=0.1$ is fit to a power law and then that function is
integrated over $0<x<0.02$.  Figure \ref{fig:combine}a shows the
combined $xF_3$ data and the corrected $F_2$ data for the four lowest
\qsq\ bins, as well as the power law fit to the low $x$ data and the
$\chi^2$ for those fits.  To be consistent with theoretical
predictions of higher twist effects on the sum rule, the $\nu$-nucleon
elastic contribution (described in \cite{bebcg}) 
was added to the integral, and both the elastic
and inelastic contributions were corrected for target mass 
effects \cite{bebcg}.
Figure \ref{fig:combine}b shows $\int F_3dx$ as a function of \qsq\ and
the theoretical prediction (see equations \ref{eqn:glsalf} and 
\ref{eqn:alfq2}) assuming $\lmsb^{(5)}=150\,MeV$.

\begin{figure}[t]
    \begin{center}
    \leavevmode   \epsfxsize=4.5in
    \epsfbox{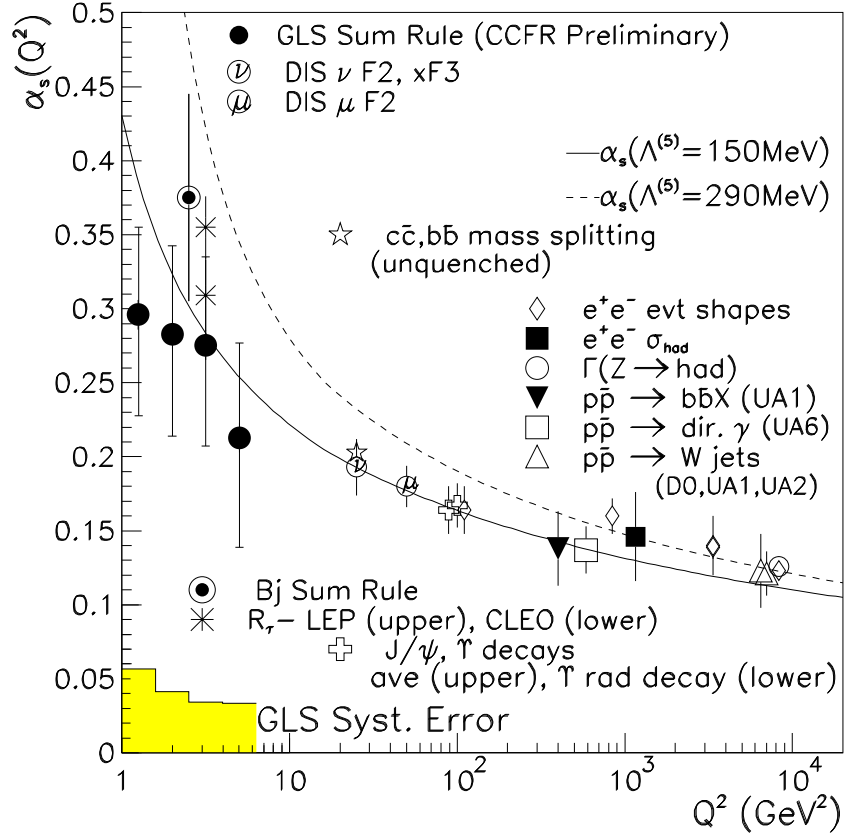}
    \end{center}
    \vspace{-.4in}  
    \caption{$\alpha_s$ as measured at several different \qsq\ values, and
             the expected \qsq\ dependence of the 3-loop (see equation 
             \ref{eqn:alfq2})  
             $\alpha_s$ for $\lmsb^{(5)}=150\,MeV$, 
             and $\lmsb^{(5)}=290\,MeV$.  Only the statistical errors on 
             the GLS points are plotted at the GLS values, 
             the systematic errors (which 
             are correlated from one \qsq\ bin to the next) are 
             shown at the bottom of the plot.}   
    \label{fig:alfworld} 
\end{figure}

One can determine \alfs(\qsq) from \glsint\ by using equation 
\ref{eqn:glsalf}.  The values of \alfs(\qsq) determined by this technique
are shown in figure \ref{fig:alfworld}.  The curves plotted in
figure \ref{fig:alfworld} show the evolution of \alfs\ as a function
of \qsq\ (see equation \ref{eqn:alfq2}), for two different values of \lmsb.
From this plot it is
clear that low \qsq\ measurements have large potential to constrain
\alfs\ not only because \alfs\ is large in this kinematic region, but
because it is changing rapidly as a function of \qsq.  However, the
higher twist uncertainty in \glsint\ is also large in this kinematic
region and is the largest single systematic error in this analysis.
Evolving the four lowest data points for \alfs\ to \mztwo, we obtain
the following value for
\alfs(\mztwo):
\begin{center} 
$\alfs(\mztwo)=0.108\pm^{.003}_{.005}$(stat) $\pm.004 $(syst)
$\pm^{.004}_{.006}$ (higher twist) 
\end{center} 

For comparison with other low \qsq\ \alfs\ measurements, this corresponds 
to $\alfs(\qsq=3.0\,GeV^2)=
0.26\pm^{.02}_{.03}$(stat) $\pm .02 $(syst) $\pm .03 $(higher twist).  
Figure \ref{fig:alfworld} puts this result in the 
context of other measurements by plotting them as a function of $Q^2$.  
In general, the low \qsq\ data systematically favor a lower
\lmsb\ than do the higher \qsq\ data.  The result from this analysis is 
consistent with 
low energy measurements of \alfs.  In particular, it is consistent with 
the CCFR determination of \alfs\, from the \qsq\ evolution of $xF_3$ and $F_2$ 
for $\qsq>15\,GeV^2$ ($\alfs(\mztwo)=0.111\pm.004$), and 
about $2\sigma$ lower than that
measured from the high \qsq\ data \cite{bethke}.  With future
experimental improvements (Fermilab NuTeV experiment) 
and improved theoretical work on higher twist corrections, this
fundamental prediction of QCD has promise for being a stringent test of 
the model.

\end{document}